# Observation of a remarkable reduction of correlation effects in $BaCr_2As_2$ by ARPES

Jayita Nayak,[1] Kai Filsinger,[1] Gerhard H. Fecher,[1] Stanislav Chadov,[1] Ján Minár,[2] Emile E. D. Rienks,[3,4] Bernd Büchner,[3,4] Jörg Fink,[1,3,4,*] and Claudia Felser[1]

[1]*Max Planck Institute for Chemical Physics of Solids, 01187 Dresden, Germany*

[2]*New Technologies Research Centre,*
*University of West Bohemia, Pilsen, Czech Republic*

[3]*Leibniz Institut für Festkörper- und Werkstoff*
*IFW Dresden, D-01171 Dresden, Germany*

[4]*Institute of Solid State Physics, Dresden University of Technology,*
*Zellescher Weg 16, D-01062 Dresden, Germany*

(Dated: January 21, 2017)

## Abstract

The superconducting phase in iron-based high-$T_c$ superconductors (FeSC) – as in other unconven- tional superconductors such as the cuprates – neighbours a magnetically ordered one in the phase diagram. This proximity hints at the importance of electron correlation eff ts in these materials, and Hund's exchange interaction has been suggested to be the dominant correlation effect in FeSCs because of their multiband nature. By this reasoning, correlation should be strongest for materi- als closest to a half filled **3d**-shell (Mn compounds, hole doped FeSCs) and decrease for systems with both higher (electron doped FeSCs) and lower (Cr-pnictides) 3d counts. Here we address the strength of correlation effect in $BaCr_2As_2$ by means of angle-resolved photoelectron spectroscopy (ARPES) and first principles calculations. This combination provides us with two handles on the strength of correlation: First, a comparison of the experimental and calculated effective masses yields the correlation induced mass renormalisation. In addition, the lifetime broadening of the experimentally observed dispersions provides another measure of the correlation strength. Both approaches reveal a reduction of electron correlation in $BaCr_2As_2$ with respect to systems with a **3d** count closer to five. Our results thereby support the theoretical predictions that Hund's exchange interaction is important in these materials.

# I. INTRODUCTION

The discovery of the high-$T_c$ iron based superconductors (FeSC)[1] has initialised strong research activities in this field.[2] Similar to the cuprates and other unconventional superconductors – in the phase diagram, temperature vs. control parameter – the superconducting domain appears at the end of an antiferromagnetic region. Therefore, it seems likely that correlation effects may be important for the FeSCs. On the other hand, different from the cuprates, the parent compounds of the FeSCs are not Mott–Hubbard insulators but they are metallic. Furthermore a multiplet analysis of X-ray absorption edges at the Fe 2p corelevel yielded values for the onsite Coulomb interaction $U$ at the iron sites smaller than 2 eV which is less than the bandwidth of the iron bands $W$ = 5 eV.[3,4] This led to the conclusion that FeSCs are moderately correlated systems.

Theoretical studies,[5–10] however, pointed out that correlation effects are important due to the multi-orbital electronic structure together with Hund's exchange interaction $J_H$. According to these studies, correlation effects should be strongest for compounds with a nearly half filled $3d$ shell (eg. Mn compounds and hole doped ferropnictides), and these effects should be smaller for compound having number of $3d$ electrons at the transition metal closer to 6 (electron doped ferropnictides), or compounds having a $3d$ count closer to 4 (e.g. chromiumpnictides).

There are several studies of the mass enhancement as a function of the number of $3d$ electrons at the transition metal in the ferropnictides that confirm these predictions.[11,12] These studies, however, suffer from a large variance of the experimental results and in the case of transport and thermal property data they cannot be assigned to particular bands. Additional angle resolved photoemission spectroscopy (ARPES) data on the scattering rate of inner hole pockets as a function of $3d$ count[13] show a strong incoherence of the charge carriers in the hole doped ferropnictides and more coherence for the electron doped compounds.

The theoretical results predicting a symmetry relative to the $3d$ count of five for a half filled $3d$ shell led to the speculation that high $T_c$ superconductivity may also appear in chromiumpnictide compounds.[14,15] To our knowledge no superconductivity was observed in BaCr$_2$As$_2$ . Mn and Cr doping of BaCr$_2$As$_2$ turned out to be detrimental to superconductivity.[16] Singh *et al.* reported that BaCr$_2$As$_2$ exhibits an enhanced renormalisation of the specific heat and stronger Cr–As covalency compared to the Fe based superconductors.[17]

Very recently, the discovery of superconductivity in $K_2Cr_2As_3$[18] and high pressure superconductivity in CrAs[19] sparked renewed interest in exploring the effect of Cr doping in iron based superconductors.

In the present contribution we study the question whether correlation effects increase or decrease for pnictide compounds with less than five $3d$ electrons at the transition metal. This is a fundamental question in correlated condensed matter physics. The answer could confirm the importance of Hund's exchange interaction $J_H$ for the coherence of the charge carriers in these compounds and it could confirm or discard the existence of a "Hund's metal".[5]

We use ARPES,[20] to yield information of the mass renormalisation, related to the real part of the self-energy $\Re\Sigma$, and on the scattering rate or the lifetime broadening $\Gamma$, which is related to the imaginary part of the self-energy $\Im\Sigma$. We point out that the $\Re\Sigma$ and the $\Im\Sigma$ are related via the Kramers–Kronig relation. We receive information on the orbital and band dependence of the strength of the correlation effects because we obtain not only energy dependent, but also momentum dependent information. The experimental results are compared to band structure calculations in the local density approximation (LDA) to density functional theory. The results on the mass renormalisation as well as the scattering rate indicates a decrease of the correlation effects for compounds with a $3d$ count smaller than five which confirms the theoretical predictions for the importance of Hund's exchange interaction for compounds containing transition metals with an electron configuration which is close to a half filled $3d$ shell.

## II. EXPERIMENTAL DETAILS

Angular resolved photoelectron spectroscopy has been performed on $BaCr_2As_2$ single crystals. The details of the single crystal synthesis, characterisation, and physical properties of the samples are described elsewhere.[21] The ARPES experiments were carried out at the UE 112–PGM2b beamline of the synchrotron facility BESSY (Berlin) using the $1^3$ ARPES end station that is equipped with a Scienta R4000 energy analyser. All measurements were performed at a temperature of 1 K. The photon energies were varied from 50 to 110 eV using both horizontal and vertical polarisations. The total energy resolution was approximately 4 meV and the angular resolution was $0.2$°. The measurements were performed on in-situ cleaved samples.

The *ab initio* calculations are based on the multiple scattering approach Korringa–Kohn–Rostoker (KKR) method) and the local density approximation (LDA) to density functional theory as implemented in the Munich SPR–KKR program package.[22,23] As a first step of our investigations we performed self-consistent calculations for the 3D bulk as well as the 2D semi-infinite surface of antiferromagnetic $BaCr_2As_2$ within the screened KKR formalism. The first principles analysis of the photoelectron spectroscopy is based on the fully relativistic one-step model of photoemission in its spin density matrix formulation,[24–26] which includes all matrix-element effects, multiple scattering in the initial and final states, and all surface related effects in the excitation process. This approach has been recently successfully applied to the calculations of ARPES spectra of FeSCs.[26] All photoemission calculations were performed using the geometry and parameters of the experiment, including photon energy and polarisation.

## III. RESULTS AND DISCUSSION

In Fig. 1 (a) and (b) we show the crystal structure and the Brillouin zone (BZ) of $BaCr_2As_2$, respectively. Further, we present in Fig. 1 (c) and (d) ARPES data of $BaCr_2As_2$ near the $k_z = 0$ plane using *s*-polarised (vertically polarised) photons with the energy h$\nu$ = 70 eV. For the calculation of the $k_z$ values we use an inner potential of $V$ = 15 eV. The ARPES data are compared to the theoretical calculations. Fig. 1 (c) shows the experimental Fermi surface map. In Fig. 1 (d) we present an energy distribution map (EDM) measured along the $S' - \Gamma - S'$ direction (parallel to $k_y$) where $S' = \frac{\pi}{a}(0,1,0)$. Three hole pockets at the centre of the BZ are visible in Fig. 1 (d). The derived Fermi wave vectors are presented in Table 1 and compared to values obtained from LDA band structure calculations which are depicted in Fig. 1 (f). In Fig. 1 (e) we show a calculation of the photoelectron intensity near the Fermi energy using the LDA together with matrix elements and final state effects (see section Methods). The Fermi surface of $BaCr_2As_2$ consists of three sheets that are a small cushion shaped, closed pocket around $\Gamma$ and two corrugated, open cylinders along the $k_z$ direction as reported in References [17, 21]. The outer cylinder exhibits a strong corrugation whereas the inner pocket and the middle cylinder are rather smooth. This is different from the Fermi surface reported for $BaFe_2As_2$ by Yin *et al.*[27] The bands forming those sheets are marked by numbers (I–III) in Fig. 1 (f). The calculated Fermi surface as well as the intensity distribution agrees well with the experimental ARPES Fermi surface and intensities. The strong corrugation of the outermost cylinder is clearly visible. According to the calculations, the band forming the inner pocket around $\Gamma$ (see Fig. 1 (d)) has predominantly Cr $3d_{xy}$ and As $p_z$ orbital character. The bands forming the middle and outer cylinders around Γ (see Fig. 1 (d)) are dominated by Cr $d_{xz/yz}$ and $d_{z^2}$ states and have less contributions from As $p_x$ and $p_y$ states. The orbital character of the inner small pocket, predicted from the electronic structure calculations, is in line with the experimentally observed strong intensity along the [010] direction (see Fig. 1 (c) and (d)) which is expected for vertically *s*-polarized photons and an odd wave function (here with $3d_{xy}$ orbital character) relative to the horizontal scattering plane, but not for horizontally *p*-polarised photons (see Fig. 2 (a) and (b)).[28]

From the ARPES data along the Σ direction ($S' - Γ - S'$, see Fig. 1 (b)) we derive the "renormalised" Fermi velocities $v_F^*$ for the three Fermi surface sheets which are listed in Table 1. These values are compared to the Fermi velocities $v_F^0$ derived from LDA band structure calculations. The calculated mass renormalization

$$m^*/m^0 = v_F^0 / v_F^*\ [29]$$

for the three bands closest to Γ are presented in Table 1, as well.

Analogous results are shown in Fig. 2 for a photon energy of h$\nu$ = 88 eV. For this photon energy, normal emission corresponds to the Z point in the BZ ($k_z = \frac{\pi}{c}$ where c is the lattice constant perpendicular to the CrAs layers). The presented EDM is now measured along the F direction (S − Z − S line, see Fig. 1 (b)). The inner small pocket, indeed, is not observed near the Z point, but only the two cylinders. This is in line with the calculated band structure (see Fig. 1 (d) and 2 (d)) where only one band crosses the Λ direction ($\overline{\Gamma Z}$) and results in a closed pocket formed by band I. The apparent elongation along $k_x$ of the inner structure that is seen in Fig. 2 (a) results most probably from a superposition of the intensities from the middle and outer cylinders (II and III), where band II has a larger intensity along $k_x$ while band III has a larger intensity along $k_y$. The experimental $k_F$ values for the bands II and III are listed again in Table 1. They nicely agree with values derived from the calculations also shown in Table 1. In that table, we also present the corresponding Fermi velocities and the mass renormalisations.

Fig. 3 shows ARPES data taken with horizontally p-polarised light using photons with energies of h$\nu$ = 70 eV (Fig. 3 (a,c)) and 88 eV (Fig. 3 (d)). The Fermi surface map for $k_z$ = 0 obtained with horizontal p-polarisation using 70 eV also exhibits the three sheets of the Fermi surface. The corresponding $k_F$ values are quite similar to that obtained with s-polarisation. The corresponding calculated Fermi surface map – shown in Fig. 3 (b) – is in excellent agreement with the experiment. In the EDM measured along the $S-Z-S$ line the small, inner pocket is also absent as expected from the band structure calculation.

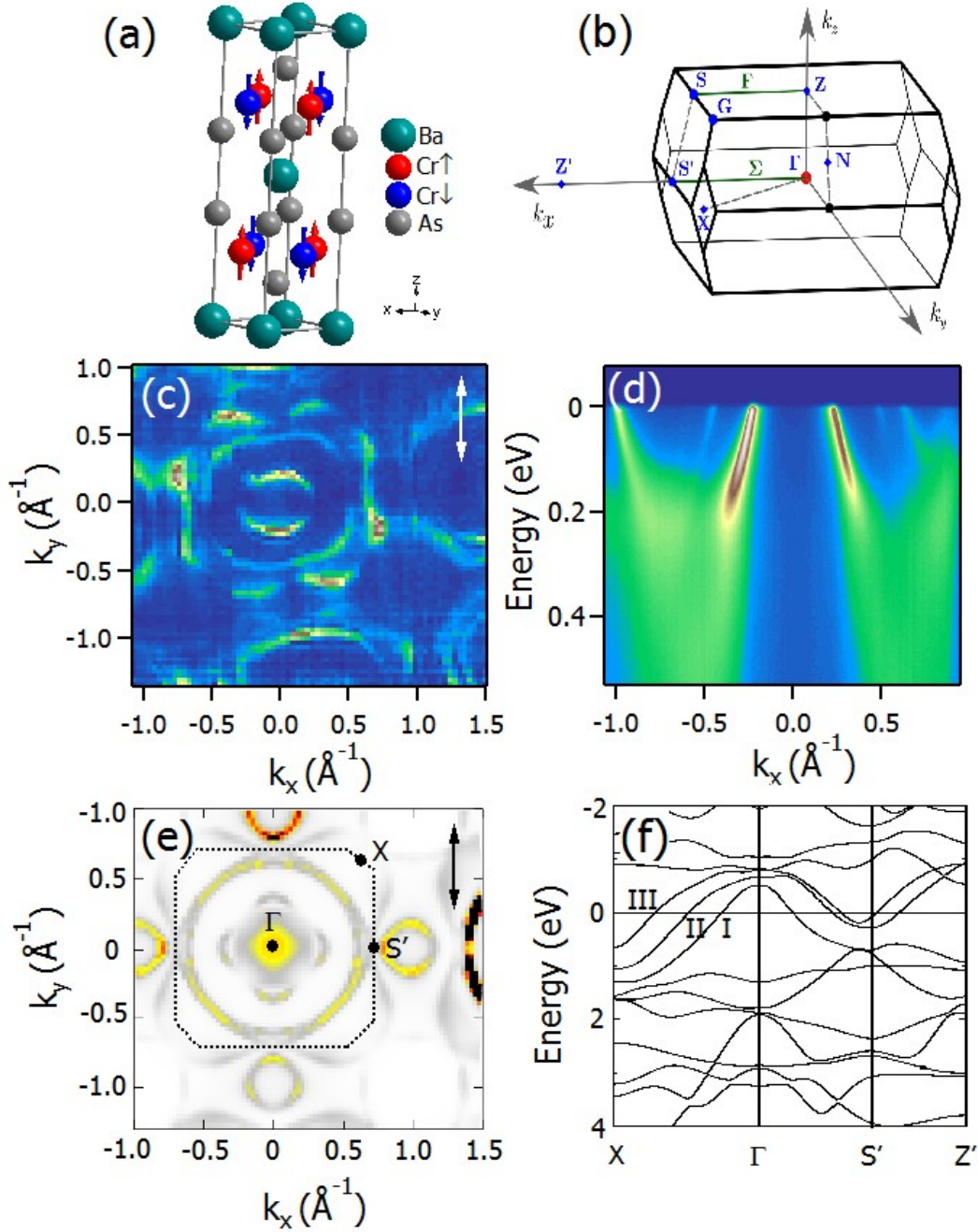

FIG. 1. (a) and (b) Crystal structure and Brillouin zone of BaCr$_2$As$_2$ . (c) and (d) ARPES Fermi surface map and EDM along $S'-\Gamma-S'$, respectively, of BaCr$_2$As$_2$ using vertically s-polarised photons with an energy of 70 eV yielding data in the $k_z$ = 0 plane. (e) Calculated photoelectron intensity at the Fermi surface. In (c) and (e) the polarisation of the photons is indicated by a double arrow. (f) Calculated band structure of BaCr$_2$As$_2$ along $X-\Gamma-S'-Z'$.

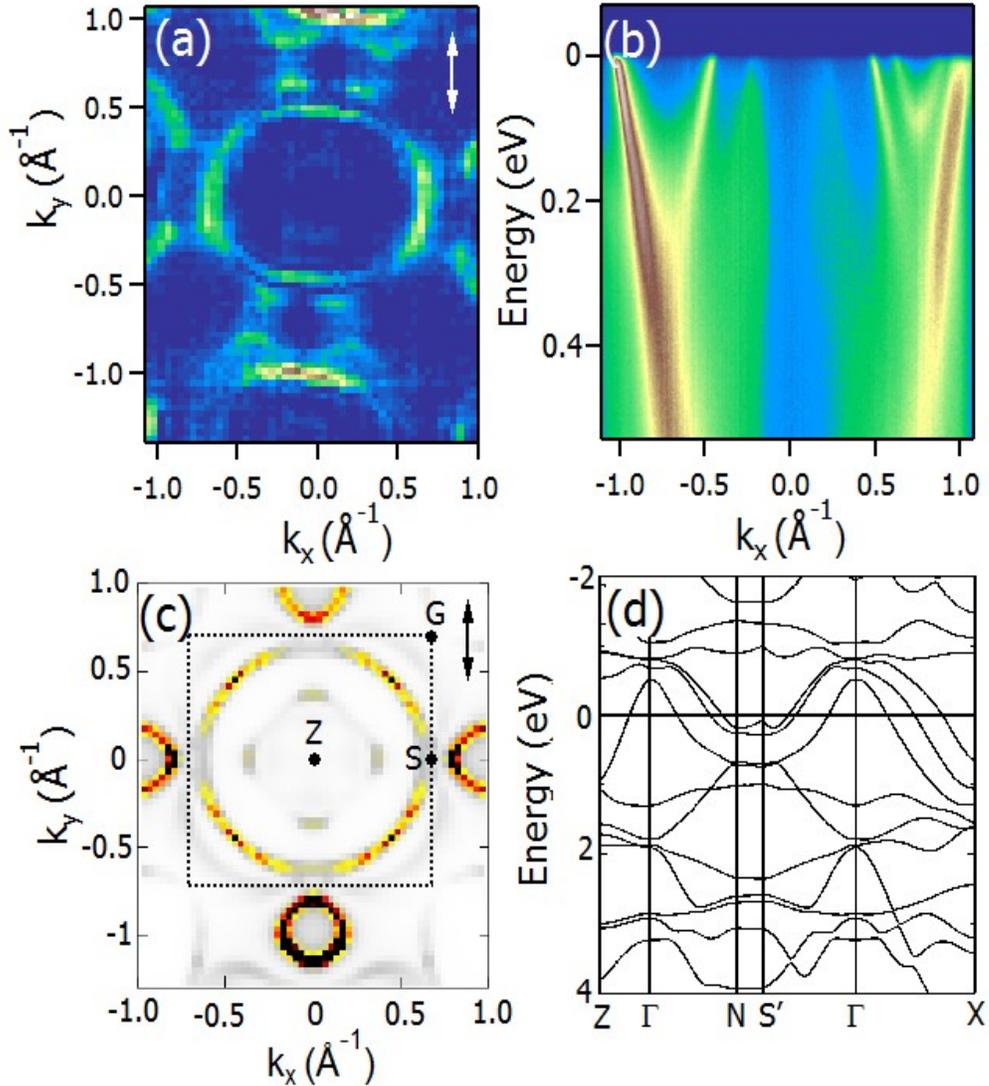

FIG. 2. Analogous data as in Fig. 1 but for a photon energy of 88 eV, yielding data in the $k_z = \frac{\pi}{c}$ plane. (a), (b) ARPES Fermi surface map and EDM along $S'-\Gamma-S'$. (c) Calculated photoelectron intensity at the Fermi surface. (d) Calculated band structure of BaCr$_2$As$_2$ along $Z-\Gamma-N-S'-\Gamma-X$.

The spectra acquired with $p$ polarised light in the $k_z = 0$ plane (see Fig. 3 (a)) and in the $k_z = \frac{\pi}{c}$ (see Fig. 3 (b)) plane exhibit the presence of additional bands below 0.15 eV that were not observed with $s$-polarisation. These bands are possibly related to Cr $3d_{z^2}$ bands. In the spectra measured along the $S'-\Gamma-S'$ line using vertically ($s$) (see Fig. 1 (c) and (d)) or horizontally ($p$) (see Fig. 3 (a) and (c)) polarised photons, all three sheets of the Fermi surface are clearly visible. This is different from the hole and electron doped ironpnictide compounds, where the intensity of the inner and the middle hole pocket changed upon changing the photon polarisation.[13,30,31] This demonstrates that in BaCr$_2$As$_2$ both, even and odd parity orbitals, contribute to each band. It indicates a stronger hybridisat between the orbitals as was already predicted in Reference [17] and it points out that the bands crossing the Fermi energy in BaCr$_2$As$_2$ are different from those forming the multisheet Fermi surface of BaFe$_2$As$_2$.

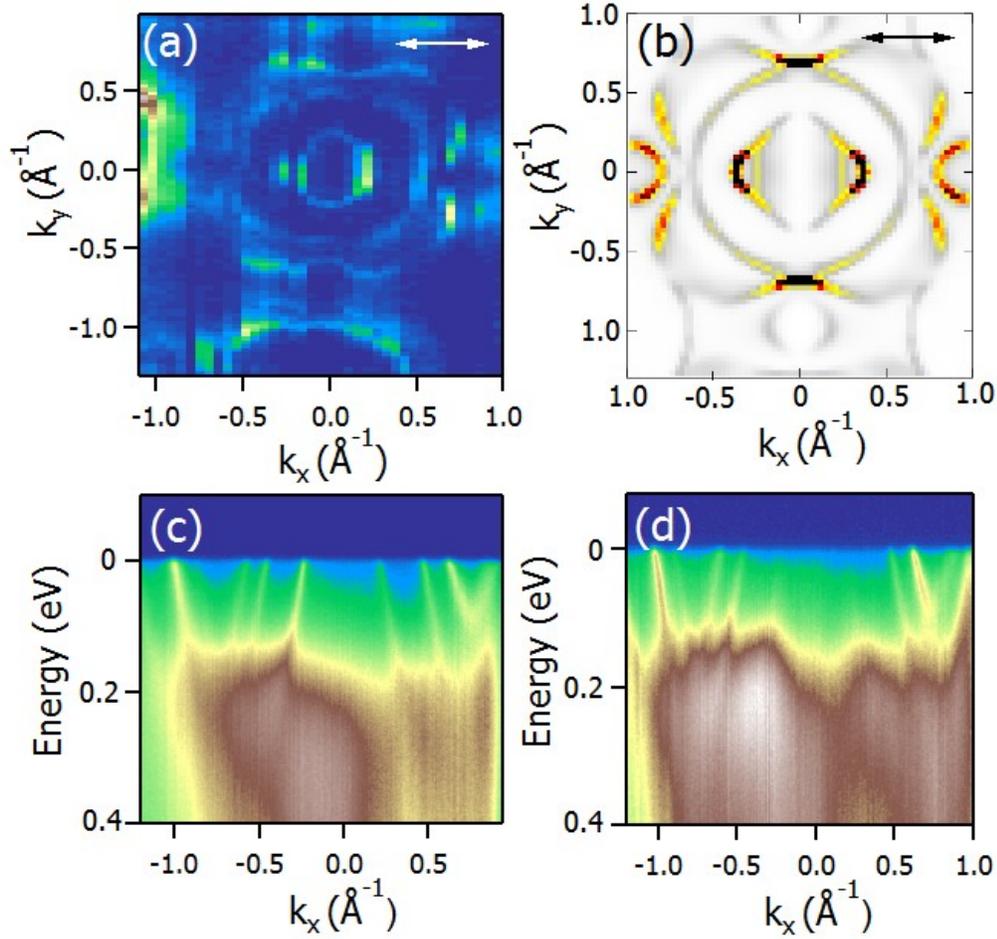

FIG. 3. (a) ARPES Fermi surface map and (b) calculated photoelectron intensity at the Fermi surface of BaCr$_2$As$_2$ for horizontally p-polarised photons with an energy of 70 eV yielding data in the $k_z = 0$ plane. (c) and (d) EDMs along $S'-\Gamma-S'$ and $S-Z-S$ recorded with photon energies of 70 eV and 88 eV, respectively. The polarisation of the photons is indicated in (a) and (b) by a double arrow.

The measured photoelectron spectra were further analyzed to extract the electron scattering rates. We performed a fitting of the momentum distribution curves by two Lorentzians for each band. The maxima of the Lorentzians, derived from constant energy cuts, result in the dispersion shown in the EDMs by the white dashed lines in Fig. 4(a) and (c) where we present ARPES data measured with *p*-polarised photons parallel to $k_y$ along $\Sigma$ and $F$ ($S'-\Gamma-S'$ and $S-\Gamma-S$), respectively. The width of the Lorentzians represents the lifetime broadening. The half width at half maximum of the momentum distribution curves multi- plied by the renormalised velocities provides the imaginary part of the self-energy $\Im\Sigma(E)$ (Fig. 4 (b,d)). In the accessible energy range, limited at low energy by the finite energy and momentum resolution and by a finite elastic scattering rate and at high energies by the presence of other bands, we observe a linear increase of the width as a function of energy. This is a hallmark of a non-Fermi-liquid regime, similar to what has been observed in the ferropnictides. In the evaluated region the imaginary part of the self-energy is approximated by $\Im\Sigma(E) = \alpha + \beta E$. $\alpha$ is determined by elastic scattering from impurities at the surface and in the bulk. Also a finite energy and momentum resolution is contributing to the ex- perimental $\alpha$ values, which are of the order of 10 to 20 meV. $\beta$ is determined by inelastic scattering processes due to electron–electron interaction.[31] The $\beta$ values derived from the ARPES data shown in Fig. 4 are presented in Table1.

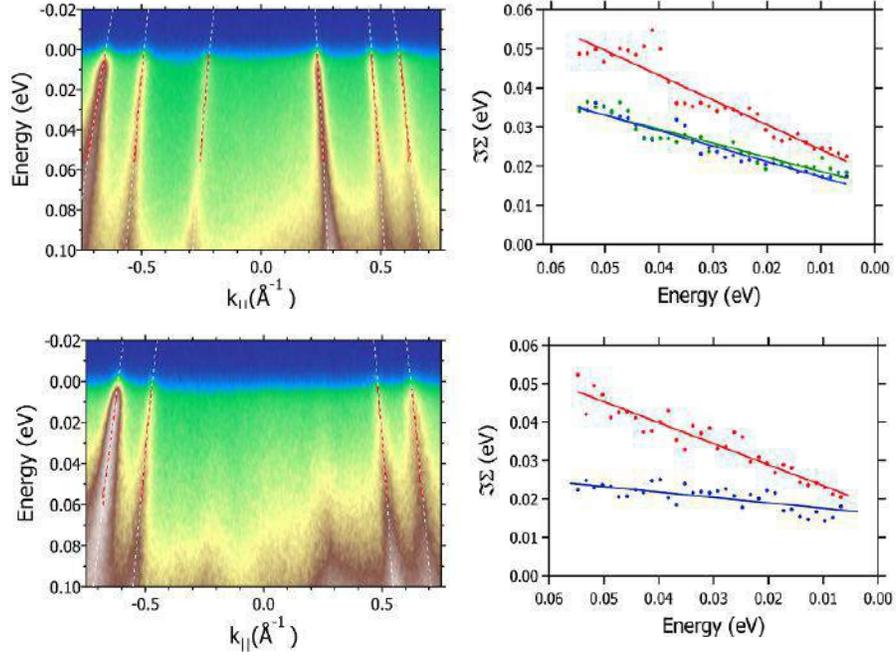

FIG. 4. EDMs and imaginary part of the self-energy $\Im\Sigma(E)$ from ARPES experiments. (a) Map along the $S'-\Gamma-S'$ direction, measured using horizontally $p$-polarised photons. (b) $\Im\Sigma$ as a function of energy for the three hole pockets together with a linear fit; green, blue, and red data correspond to band I, II, and III, respectively. (c,d) Similar as (a,b) but acquired along the $S-Z-S$ direction.

## V. DISCUSSIONS

In the electron and hole doped ferrpnictides and in ferrochalcogenides, not only a mass renormalisation has been derived from ARPES spectra, but also a band shift has been observed which lead to a correlation induced complex change of the Fermi wave vectors relative to Fermi wave vectors derived from band structure calculations.[32] Theoretical work on this phenomenon has been published in References [5, 33, 34]. The proximity of the ARPES Fermi wave vectors in $BaCr_2As_2$ to the calculated ones is a first indication that correlation effects in $BaCr_2As_2$ are smaller than in the iron based superconductors. This is supported also by the observation that different from the more correlated ferropnictides, in $BaCr_2As_2$ the $3d_{xy}$ band has the highest binding energy: in the former the $3d_{xy}$ band is up-shifted due to correlation effects, while the Fe $3d_{xz/yz}$ bands are selectively down-shifted.[34]

We compare the mass renormalisation observed in the present study on $BaCr_2As_2$ with those derived from ARPES, quantum oscillations, and thermal properties on the Fe based compounds.[11] In the latter compounds mass renormalisation values between 2 and 4 have been reported for most of the ferropnictides but in some studies values up to 18 have been published.[11] The values derived in the present study for $BaCr_2As_2$ are close to one, indicating a rather small mass renormalisation due to correlation effects mediated by electron–electron interaction. In this context we mention a recent study of the band renormalisation in $BaMn_2As_2$ and $BaMn_2Sb_2$ in which also a negligible band structure renormalisation has been observed.[35]

Next we discuss the derived scattering rates for the charge carriers in $BaCr_2As_2$ . The highest $\beta$ values in electron doped and P substituted ferropnictides are about 0.9[31] while in the hole doped compounds $\beta$ amounts to a value of 1.7.[13] In $BaCr_2As_2$ the highest value $\beta = 0.63$ is considerably reduced compared to the Fe compounds. Since the scattering rate in a local approximation is related to the effective onsite Coulomb interaction $U_{eff}$,[13,36] the reduced scattering rates could signal a reduced $U_{eff}$ and thus weaker correlation effects in the Cr based compound.

In the following we discuss the $\beta$ values of the individual bands in more detail. The $\beta$ values for the inner band I – with a predominant Cr $3d_{xy}$ orbital character – both at the $\Gamma$ and at the $Z$ point, are smaller than those derived from the other bands which have predominatly $3d_{xz/yz}$ orbital character. This is in line with calculations of the electronic susceptibility in ferropnictides in the framework of combined density functional dynamical mean field theory (DFT+DMFT) yielding higher scattering rates for the $3d_{xz/yz}$ than for the $3d_{xy}$ bands.[34] It is interesting that in BaCr$_2$As$_2$ the outer $3d_{xz/yz}$ band III with the highest $k_F$ value has the highest scattering rate while in the ferropnictide the highest scattering rates appear in the inner hole pocket with the smallest $k_F$ value.

One may think that the reduced correlation effects in BaCr$_2$As$_2$ may be related to the antiferromagnetic order of this compound. However, in a recent study of Mn compounds[35] it was pointed out that in BaFe$_2$As$_2$ the strength of the correlation effects does not change when moving from the paramagnetic to the antiferromagnetic phase.[37] This statement is in line with DFT+DMFT calculations for the mass renormalisation in BaFe$_2$As$_2$ which predicted only a minor reduction when going from the paramagnetic to the antiferromagnetic state.[38]

The reduction of the correlation effects in BaCr$_2$As$_2$ relative to the hole doped ferropnictides, derived from the real (related to the mass enhancement) and the imaginary part (related to the scattering rate) of the self-energy, perfectly agrees with the predictions derived from DFT+DMFT calculations and the conception of a "Hund's metal" that leads to strong scattering rates connected with incoherence for the half filled $3d$ shell. There, $U_{eff}$ is enhanced due to the Hund's exchange interaction $J_H$. On the other hand when moving away from half filling to lower or higher $3d$ counts, $U_{eff}$ is reduced by the Hund's exchange interaction according to these calculations. This would lead to reduced correlation effects and more coherent quasiparticles for smaller or larger filling of the $3d$ shell. The present study on the Cr compound showing a reduced scattering rate and a negligible band renormalisation when compared to hole doped ferropnictides confirms the proposed theoretical concept of Hund's metals.

**Table 1, Fermi wave vectors $k_F$[$^{-1}$], Fermi velocities $v_F$[eV], mass renormalisations $m^*/m^0$, and slopes $\beta$ of the imaginary part of the self-energy for the hole pockets in BaCr$_2$As$_2$. (See Fig. 1 (b) for the directions in the Brillouin zone.)**

| Fermi surface sheet | inner (I) | middle (II) | outer (III) |
|---|---|---|---|
| $k_F$ from ARPES along $S'-\Gamma-S'$ | 0.23±0.05 | 0.48±0.05 | 0.61±0.05 |
| $k_F$ from LDA along $S'-\Gamma-S'$ | 0.25 | 0.55 | 0.62 |
| $k_F$ from ARPES along $S-Z-S$ | - | 0.48±0.05 | 0.62±0.05 |
| $k_F$ from LDA along $S-Z-S$ | - | 0.53 | 0.68 |
| $v_F$ from ARPES along $S'-\Gamma-S'$ | 1.65±0.10 | 1.47±0.10 | 0.84±0.10 |
| $v_F$ from LDA along $S'-\Gamma-S'$ | 1.33 | 1.03 | 1.01 |
| $v_F$ from ARPES along $S-Z-S$ | - | 1.03±0.10 | 1.00±0.10 |
| $v_F$ from LDA along $S-Z-S$ | - | 1.06 | 1.08 |
| $m^*/m^0$ along $S'-\Gamma-S'$ | 0.81±0.05 | 0.70±0.05 | 1.20±0.14 |
| $m^*/m^0$ along $S-Z-S$ | - | 1.03±0.10 | 1.08±0.10) |
| $\beta$ along $S'-\Gamma-S'$ | 0.35±0.10 | 0.39±0.10 | 0.63±0.10 |
| $\beta$ along $S-Z-S$ | - | 0.14±0.10 | 0.54±0.10 |

**ACKNOWLEDGMENTS**

We acknowledge Bessy for providing experimental time with proposal No 16103260. J.F. and B.B. acknowledge support by the German Research Foundation, the DFG, through the priority program SPP 1458. J.M. acknowledges the support by CEDAMNF (CZ02.01.01/0.0/0.0/15_003/0000358) of The Ministry of Education, Youth and Sports (Czech Republic).

---

* j.fink@ifw-dresden.de